\documentclass[10pt,twocolumn]{article}
\usepackage{multicol}
\usepackage{graphicx}
\usepackage{amsfonts}
\usepackage{amsmath}
\usepackage{amssymb}
\usepackage{mathrsfs} 
\usepackage{lipsum}
\usepackage{color}
\usepackage[squaren,Gray,cdot]{SIunits}
\usepackage[top=2cm, bottom=2cm, left=1.5cm, right=1.5cm]{geometry}
\usepackage{mathtools, cuted}
\usepackage{array,amsmath,booktabs}
\usepackage[font=small, labelfont={bf,normalsize,sf}]{caption}
\usepackage{caption}
\usepackage{subcaption}
\usepackage[hyphens]{url}
\usepackage{multirow}

\usepackage{authblk}
\usepackage{abstract}

\usepackage{sectsty}
\allsectionsfont{\sf}

\usepackage{appendix}

\newcommand{\mNperm}{\milli\newton\per\meter}
\newcommand{\mPas}{\milli\pascal\cdot\second}
\newcommand{\Spread}{|\mathcal{S}|}

\title{\sf Interfacial Dripping Faucet: Generating Monodisperse Liquid Lenses}

\author[1,2]{Lorène Champougny}
\author[3]{Vincent Bertin}
\author[3]{Jacco H. Snoeijer}
\author[1,2]{Javier Rodr\'{i}guez-Rodr\'{i}guez}

\affil[1]{Carlos III University of Madrid, Thermal and Fluids Engineering Department, Avenida de la Universidad, 30 (Sabatini building), 28911 Leganés (Madrid), Spain}
\affil[2]{Carlos III University of Madrid, "Gregorio Mill\'{a}n Barbany" University Institute, Avenida de la Universidad, 30 (Sabatini building), 28911 Leganés (Madrid), Spain}
\affil[3]{Physics of Fluids Group, Faculty of Science and Technology, University of Twente, 7500 AE Enschede, The Netherlands}
\date{}
\begin{document}

%
\twocolumn[
\begin{@twocolumnfalse}
\maketitle       
\begin{abstract}
\vspace{-12mm}
We present a surface analog to a dripping faucet, where a viscous liquid slides down an immiscible meniscus. 
Periodic pinch-off of the dripping filament is observed, generating a succession of monodisperse floating lenses.
We show that this interfacial dripping faucet can be described analogously to its single-phase counterpart, replacing surface tension by the spreading coefficient, and even undergoes a transition to a jetting regime. 
This liquid-liquid-gas system opens perspectives for the study of the dynamics of emulsions at interfaces.
\end{abstract}
\vspace{6mm}
\end{@twocolumnfalse}]
%
\section*{Introduction} \label{sec:intro}
%
Since the times of J. C. Maxwell and Lord Rayleigh, the dripping faucet has attracted the attention of fluid mechanicists, applied mathematicians and material scientists alike.
Not only did it turn out to be a rich dynamical system -- exhibiting behaviors such as period doubling, chaos, and dripping-to-jetting transition \cite{Subramani2006, Clanet1999} -- but also a very fruitful setting to study the fundamentals of liquid fragmentation \cite{Eggers2008}.
In practice, the dripping faucet has served as a conceptual basis to develop tools for the generation of controlled liquid dispersions, either in air or in an immiscible liquid (emulsion).
Techniques based on the dripping faucet are used in the pharmaceutical industry, micro- and nanotechnology, or metallurgy to produce monodisperse droplets with sizes ranging from several microns to around the millimeter \cite{Montanero2020}.

Three-fluid dispersions (liquid-liquid-gas) are a common occurrence in food \cite{Hotrum2004} and cosmetic products \cite{Venkataramani2020}, coatings \cite{Purohit2014}, as well as in separation \cite{Su2006, Rossen2017} and cleaning processes \cite{Shaw2003, Nissanka2018, Gupta2017}.
In principle, techniques inspired from the dripping faucet can be applied to make dispersions involving more than two fluid phases -- for example multiple emulsions for encapsulation \cite{Chu2007}, in which phases meet two by two. 
However, as soon as more than two immiscible phases meet, a triple contact line appears, thereby introducing a new ingredient with its own physics.
Despite their practical interest, associated three-phase fluid systems are still sparsely understood.
Recent advances include the shape of droplets sitting at the interface between an immiscible liquid and a gas \cite{Sebilleau2013, Ravazzoli2020} -- commonly referred to as {\em liquid lenses} \cite{Johnson1985, Nepomnyashchy2021} -- and their coalescence dynamics \cite{Hack2020, Tewes2021, Scheel2023}.

In this Letter we present a system analogous to the dripping faucet, where a dispersed phase made of liquid 2 is generated directly at the surface of an immiscible, denser liquid 1.
We show that this original system, dubbed ``interfacial dripping faucet'', is able to periodically generate monodisperse liquid lenses in a controlled fashion, and we rationalize the volume of the produced lenses.
This configuration provides a simple route for the generation of large collections of liquid lenses (emulsions at interfaces) but also offers a controlled setting to investigate pinch-off dynamics in the presence of a liquid-liquid-gas contact line.
%
\section*{System description} \label{sec:syst_description}
%
\begin{figure}
    \centering
    \includegraphics[width =\linewidth]{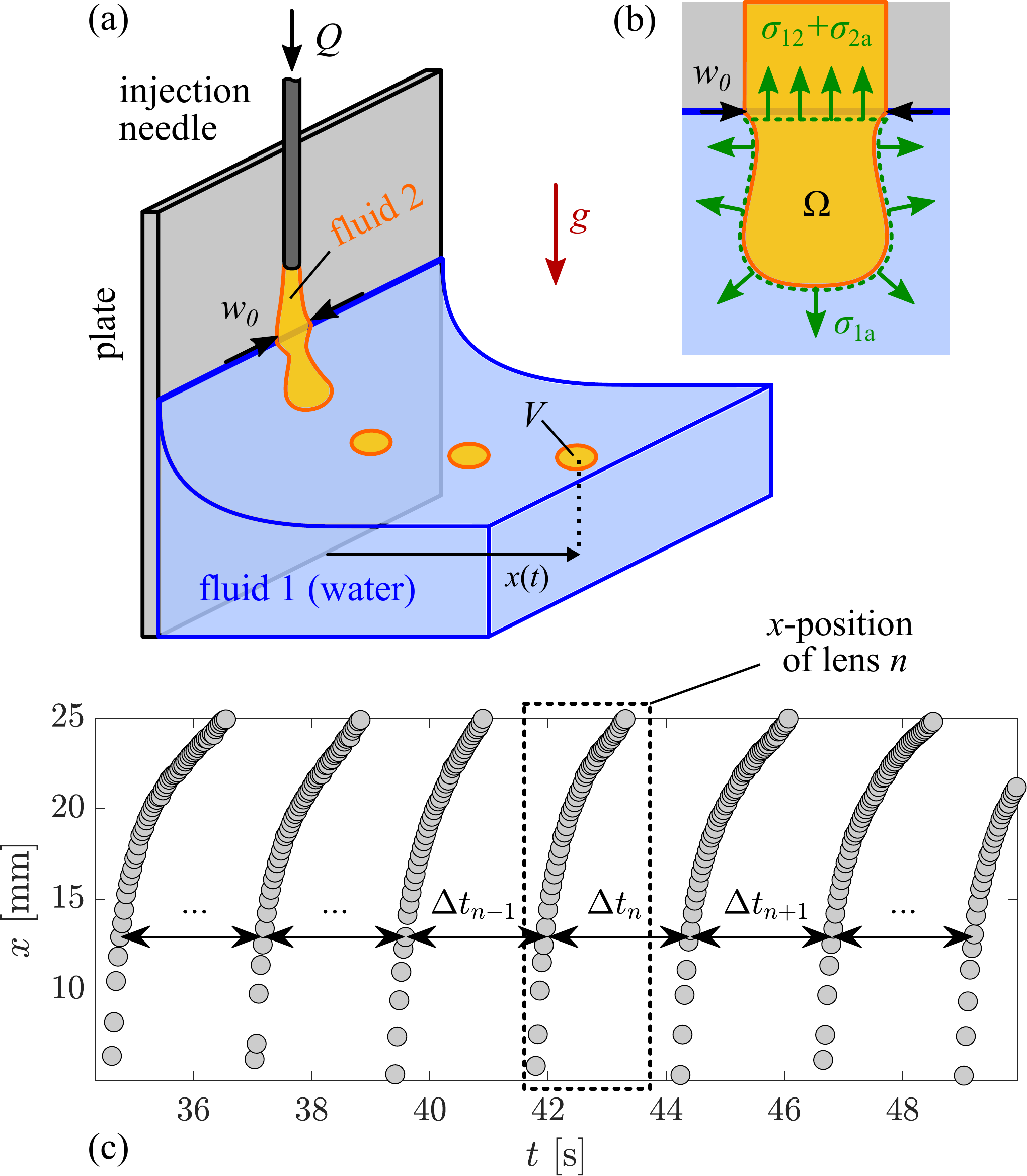}
    \caption{(a) Geometry of the interfacial dripping faucet. (b) Capillary forces exerted on a control volume $\Omega$ of liquid 2, as it enters the meniscus region (blue). (c) Example of measured time evolution of the $x$ position for consecutive lenses in the periodic pinch-off regime, allowing us to compute the dripping intervals $\Delta t_n$.}
    \label{fig:sketch_IDF_geometry}
\end{figure}
%
%
The interfacial dripping faucet geometry is sketched in Fig.~\ref{fig:sketch_IDF_geometry} (see Fig.~\ref{fig:setup_picture} in Appendix for an experimental picture).
A perfectly wetting vertical plate is partially dipped into a bath of liquid 1, which forms a meniscus. 
An injection needle placed vertically against the plate injects liquid 2 at a constant flow rate $Q$.
Liquid 2 forms a rivulet flowing down the dry substrate until it meets the three-phase contact line between the plate and the meniscus of liquid 1.
Downstream, a hanging ligament of liquid 2 forms at the surface of liquid 1. 
The fate of this ligament, of width $w_0$ at the contact line, depends on the liquid properties and flow conditions.

%
In our experiments, liquid 1 is ultrapure water, with density $\rho_1 = 998~\kilo\gram\per\cubic\meter$, dynamic viscosity $\mu_1 = 0.98~\mPas$ and surface tension $\sigma_{1a} = 72~\mNperm$ in our room conditions ($T = 21~\degreecelsius$).
Six different alkanes and mineral oils, all lighter than water, are used as the dispersed phase (liquid 2).
These liquids have similar densities $\rho_2$, surface tensions $\sigma_{2a}$ and interfacial tensions with water $\sigma_{12}$, but dynamic viscosities $\mu_2$ spanning almost four decades ($1.5~\mPas$ for dodecane to $7571~\mPas$ for mineral oil RTM30).
The spreading parameter $\mathcal{S} = \sigma_{1a} - (\sigma_{12} + \sigma_{2a})$ of all those liquid pairs fulfills the condition $ - 2 \, \mathrm{min}(\sigma_{2a}, \sigma_{12}) < \mathcal{S} < 0$, meaning that liquid 2 is able to form stable and uncloaked lenses on water \cite{Ravazzoli2020} (see also section~\ref{apdx:matmet} in Supp. Mat.).

\begin{figure}
    \centering
    \includegraphics[width = \linewidth]{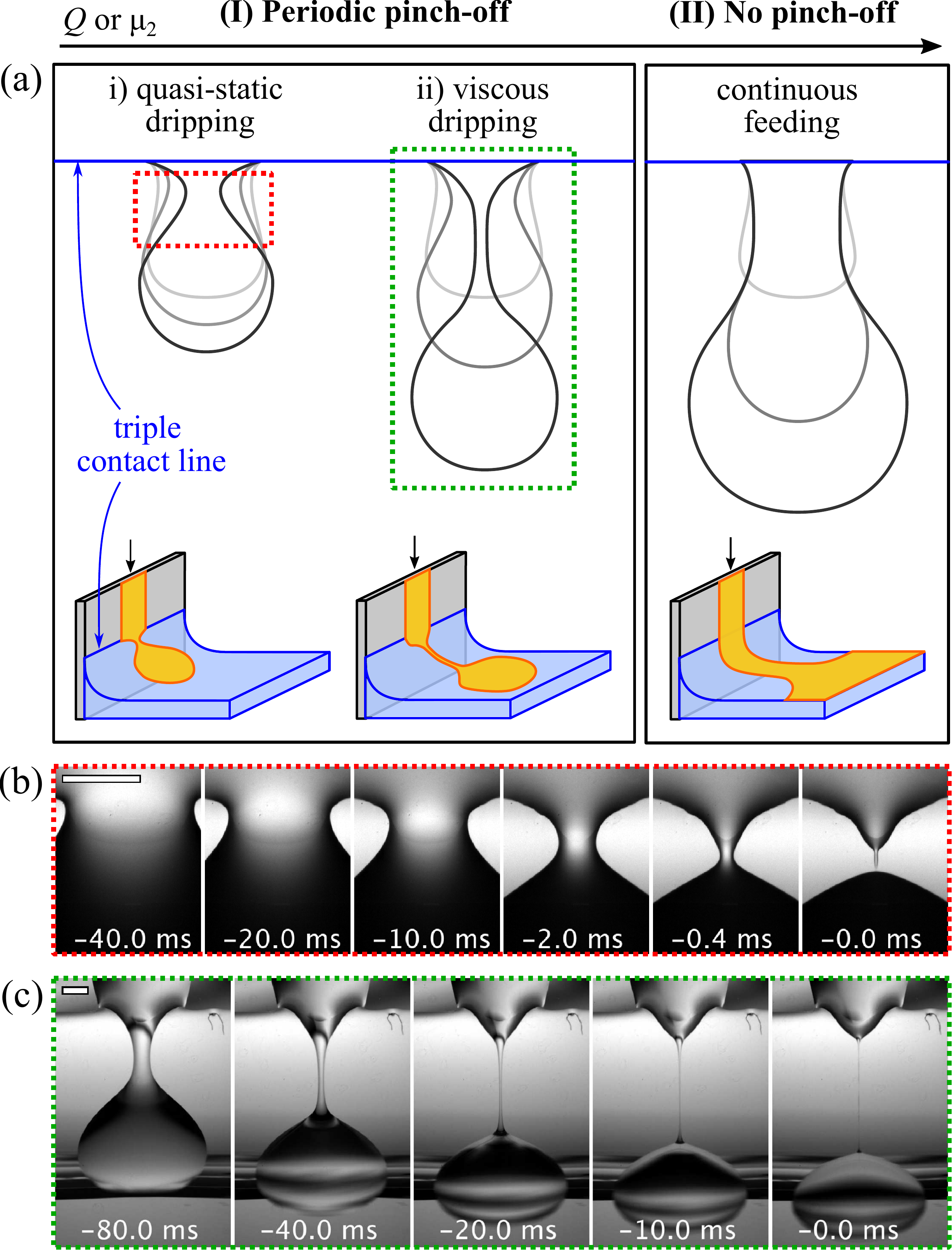}
    \caption{(a) Regimes of the interfacial dripping faucet. The top sketches qualitatively represent the contour of the hanging ligament (front view) at different times, increasing from lighter to darker shade. The bottom sketches show a side view of the hanging ligament on the water meniscus. (b) Experimental time sequence of pinch-off for dodecane with $Q = 5~\micro\liter\per\minute$ (quasistatic dripping, i). (c) Time sequence of pinch-off for S200 mineral oil, $Q = 50~\micro\liter\per\minute$ (viscous dripping, ii). Both scale bars are $500~\micro\meter$.}
    \label{fig:regime_overview}
\end{figure}

%
Depending on the viscosity $\mu_2$ and the injection flow rate $Q$ of the dispersed phase, two regimes can be observed, as sketched in Fig.~\ref{fig:regime_overview}(a).
In the first regime (I), found for small values of $Q$ and/or $\mu_2$, the hanging ligament of liquid 2 pinches off periodically, leading to the formation of identical liquid lenses that ``drip'' along the water meniscus (Movie S1).
This periodic dripping regime remains stable as long as the injection continues, and the water surface is not entirely covered with lenses.
High-speed imaging reveals that two limiting cases can be distinguished, depending on the pinch-off location and morphology.
Quasistatic dripping (i) is characterized by a necking and pinch-off close to the contact line (Fig.~\ref{fig:regime_overview}b, Movie S2).
In viscous dripping (ii), the hanging ligament stretches into a long thread that eventually pinches off further down on the meniscus (Fig.~\ref{fig:regime_overview}(c), Movie S3).
Increasing the flow rate $Q$ and the viscosity $\mu_2$ of liquid 2, another regime (II) emerges, in which the ligament never breaks (Movie S5).

\section*{Periodic pinch-off regime}
%
We first focus on characterizing the periodic pinch-off regime (I).
Experimentally we acquire videos of the meniscus region, where lenses are formed and break up from the hanging ligament. 
Once a lens has pinched off, its position $x$ in the direction perpendicular to the plate is recorded as a function of time $t$ using an automated in-house image processing code.

%
Fig.~\ref{fig:sketch_IDF_geometry}(c) displays examples of $x$--$t$ trajectories of a series of lenses detached consecutively, here for dodecane injected at $Q = 50~\micro\liter\per\minute$.
The time at which the lens reaches a reference position ($\sim 13~\milli\meter$ in this case) allows us to robustly measure the time interval $\Delta t_n$ between lens $n$ and the following lens $n+1$, referred to as the dripping interval.
For a given liquid and flow rate, the average value $\Delta t$ of the dripping intervals will be referred to as the dripping period (see section~\ref{apdx:image_processing} in Supp. Mat.). In all our experiments, the typical variability in dripping interval is $\lesssim 10~\%$, comparable to that of the classical dripping faucet in the constant-dripping interval regime \cite{Subramani2006}. Moreover, we do not observe any trend in these small fluctuations of $\Delta t_n$, which seem to be stochastic in nature (see Fig.~\ref{fig:dt_vs_Q} in Supp. Mat.). We therefore treat the dripping period $\Delta t$ as a well-defined observable for a given set of flow parameters.

%
\begin{figure}
    \centering
    \includegraphics[width=\columnwidth]{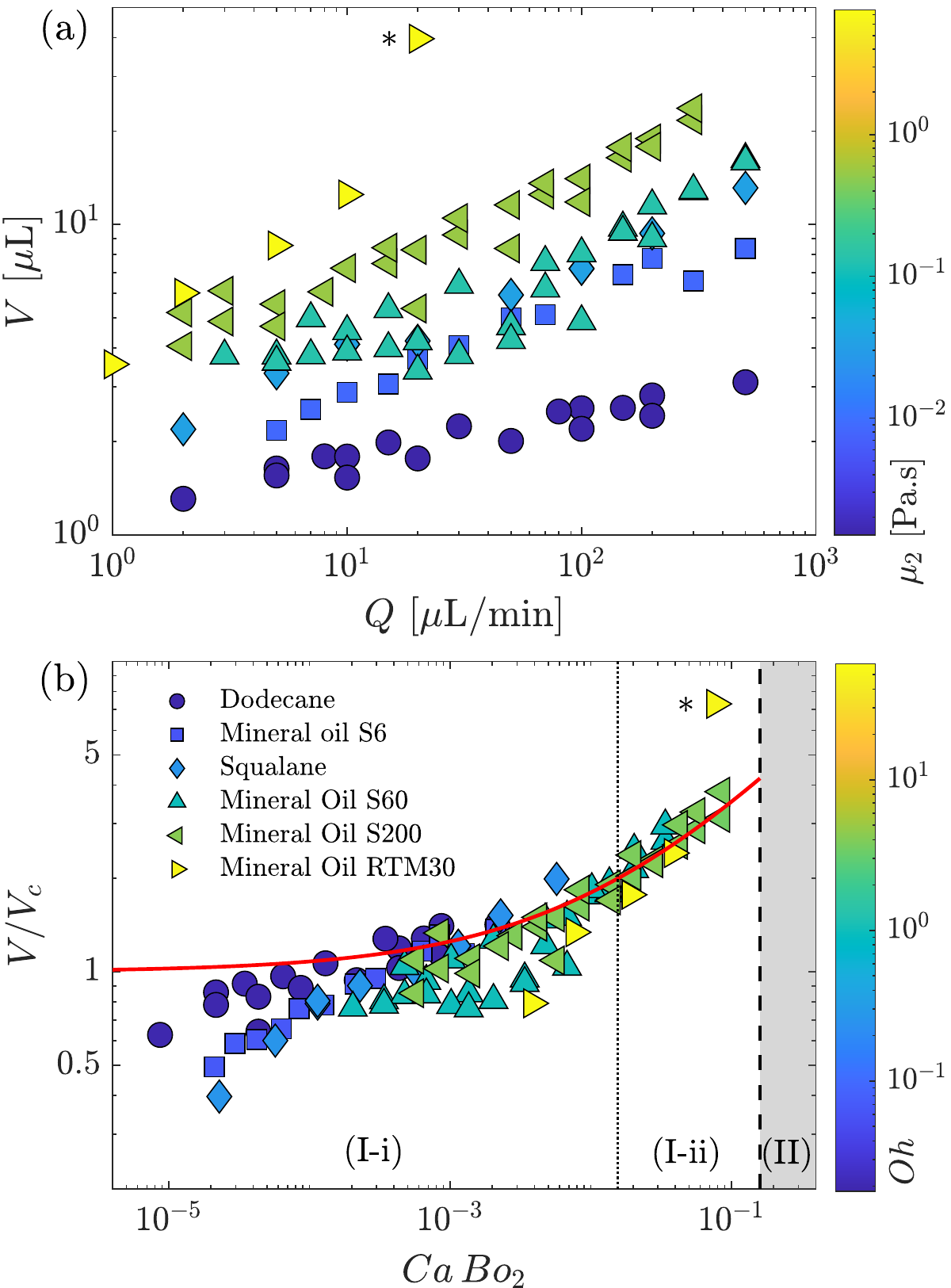}
    \caption{(a) Average liquid lens volume, $V = Q \, \Delta t$, as a function of the flow rate $Q$. The color encodes the liquid dynamic viscosity $\mu_2$. (b) Dimensionless lens volume $V/V_\mathrm{c}$ as a function of the dimensionless group $\mathrm{Ca}\,\mathrm{Bo}_2$. The color encodes the Ohnesorge number $Oh = \mu_2 / \sqrt{\rho_2 g w_0}$. The solid line corresponds to equation (\ref{eq:scaling_dimensionless}) with $C = 8$. The dotted and dashed lines show the transitions between periodic behaviors (i) and (ii), and between regimes (I) and (II), respectively. The asterisk highlights an experimental point where we observed that the ligament wobbles several times before eventually undergoing pinch-off.}
    \label{fig:Vadim_vs_Ca}
\end{figure}
When dripping is periodic, the volume $V$ of the emitted lenses and the dripping period $\Delta t$ are related by mass conservation, $Q = V\Delta t$. 
This relation holds as long as the volume of any satellite lens created at pinch-off can be neglected. 
Using high-speed recordings of the pinch-off, we checked that the satellites are orders of magnitude smaller than the main lens, consistently with the observations of Ref.~\cite{Burton2007}.
Fig. \ref{fig:Vadim_vs_Ca}(a) shows the lens volume $V$, deduced from the dripping period $\Delta t$, as a function of the flow rate $Q$ for various injected liquids, color-coded according to their viscosity.
We observe that the volume grows slowly with the flow rate for a given liquid, with a more pronounced slope the more viscous the liquid is. 
For a $250$-fold increase in flow rate $Q$, the lens volume $V$ grows by a factor $3$ in the case of the least viscous liquid we studied (dodecane), and by a factor $5$ for the much more viscous S200 mineral oil.
To rationalize this behavior we propose a mechanistic model based on two key processes that must be completed for a lens to detach. First, the injected liquid accumulates at the plate-water-gas contact line until a critical volume is reached, where capillary forces are no longer able to balance the weight of the hanging ligament. Then, liquid starts to drip and a second process starts: the ligament is stretched by a balance between gravity and elongational viscous stresses, until pinch-off eventually occurs. Let us now estimate the timescales of those two processes separately.

%
Analogously to pendant drops \cite{DeGennes2004}, there is a maximum volume $V_c$ of liquid 2 that capillary forces can sustain against gravity in the interfacial dripping faucet.
To determine $V_c$, we apply a force balance on a control volume $\Omega$ containing the injected liquid at the moment of lens detachment, highlighted by a dashed line in Fig.~\ref{fig:sketch_IDF_geometry}(b).
The total capillary force exerted on $\Omega$ is $F_\mathrm{c} \sim w_0 (\sigma_{1a} - \sigma_{2a} - \sigma_{12}) = w_0 \mathcal{S}$ (see section~\ref{apdx:ligament_modelling} in Supp. Mat.). 
Balancing $F_c$ with the liquid weight in the hanging filament, $\rho_2 g V_c$ (where $g = 9.8~\meter\per\second^2$ is the acceleration of gravity) we find the critical volume $V_c \sim w_0 \Spread / \rho_2 g$. 
In the limit where the flow rate $Q$ and/or viscosity $\mu_2$ are small enough (limit (i) in Fig.~\ref{fig:regime_overview}), $V_c$ is expected to be a good estimate for the final detached lens volume. 
We thus denote the corresponding dripping period $\Delta t^{\mathrm{(i)}} = V_c/Q$.
Note that, to compute $V_c$, we treated the water/air surface -- and thus the liquid ligament -- as parallel to gravity. This assumption is valid as long as pinch-off occurs in the vicinity of the plate/water/air triple contact line, where the water meniscus is perfectly vertical thanks to the hydrophilic treatment applied to the plate. In our experiments, the typical lens size $V^{1/3}$ is smaller than the water meniscus extension, of order $\ell_{c1} = \sqrt{\sigma_{1a}/\rho_1 g}$, supporting the assumption of a vertical ligament ($0.4 \lesssim V^{1/3}/\ell_{c1} \lesssim 1$). Additionally, because the free surface of the meniscus is nearly vertical in the region where lenses form, buoyancy forces are approximately horizontal there (see section~\ref{apdx:ligament_modelling} in Supp. Mat.), as these forces must be perpendicular to the water/air interface. This justifies neglecting buoyancy in the vertical force balance.

%
Once the critical volume is reached, the hanging ligament starts stretching under the action of gravity, with its motion resisted by viscous forces.
Friction due to the water substrate can be neglected as $\mu_1 \ll \mu_2$ in most of our experiments, and the shear strain rate in the bath (of size $\sim 10~\centi\meter$) is much smaller than the longitudinal strain rate inside the ligament (of millimetric length scale $w_0$).
Viscous stresses are then dominated by the Trouton stresses inside the ligament, like in the viscous round jet \cite{Eggers1994}, rather than by the shear stresses exerted by the bath on the ligament. 
Exact solutions have been obtained for the classical, axisymmetric dripping faucet \cite{Wilson1988}, but we focus here on scaling arguments.
Suppose that a viscous liquid starts to be injected at $t=0$ from a nozzle of cross-sectional area $A_0$. 
At time $t$, the weight of liquid hanging is $\rho_2 g Q t$, which induces a stretching of the ligament that is mediated by viscosity. 
The elongational viscous stress is $3\mu_2 \dot{\epsilon}$, where $\dot{\epsilon}$ is the stretching rate, and $3\mu_2$ the Trouton viscosity \cite{Howell1994}.
The corresponding viscous force in the ligament is $3\mu_2 \dot{\epsilon} A \sim \mu_2 \partial A/\partial t$, where $A$ is the instantaneous ligament cross section. 
Balancing this force with the lens weight, we find the typical stretching time of the ligament, $(\mu_2 A_0 / \rho_2 g Q)^{1/2}$.
For a large enough flow rate $Q$ and/or viscosity $\mu_2$ (limit (ii) in Fig.~\ref{fig:regime_overview}), we expect this stretching process to take much longer than reaching the critical volume $V_c$, therefore setting the dripping period, $\Delta t^{\mathrm{(ii)}} = (\mu_2 A_0 / \rho_2 g Q)^{1/2}$.

%
Most experiments actually lie between the two limits discussed previously. 
Since the quasistatic filling and viscous stretching steps occur in a mostly sequential manner (stretching starts only once $V_c$ is reached), we propose to approximate the dripping period in the general case by the sum of the time taken by those two processes: $\Delta t = \Delta t^{\mathrm{(i)}} + C \Delta t^{\mathrm{(ii)}}$, with $C$ a dimensionless numerical constant. This constant encapsulates the effect of the complex geometry of the ligament and cannot be determined solely by scaling arguments (see section~\ref{apdx:ligament_modelling} in Supp. Mat.). Such a constant is also needed in the equations used to predict the drop volume in the classical dripping faucet \cite{Clift2005}. Multiplying the previous expression by $Q / V_\mathrm{c}$, we arrive at 
\begin{equation}
    \frac{V}{V_\mathrm{c}} = 1 + C \, \left(\mathrm{Ca} \, \mathrm{Bo}_2\right)^{1/2}.
    \label{eq:scaling_dimensionless}
\end{equation}
We have introduced here the Bond number $\mathrm{Bo}_2 = \rho_2 g w_0^2/ \Spread$ of the hanging ligament, and the capillary number $\mathrm{Ca} = \mu_2 U / \Spread$ based on the characteristic injection speed at the contact line, $U=Q/A_0$. 
The cross-sectional area $A_0$ of the ligament is deduced from the experimental value of the ligament width $w_0$ and the wetting properties.
We show in section~\ref{apdx:ligament_modelling} in the Supplemental Material that the interpolation given by Eq.~\eqref{eq:scaling_dimensionless} is a good approximation to the exact solution derived by Wilson \cite{Wilson1988} for the vertical, axisymmetric dripping faucet.

Fig. \ref{fig:Vadim_vs_Ca}(b) shows the dimensionless volume $V/V_\mathrm{c}$ as a function of the dimensionless group $\mathrm{Ca}\,\mathrm{Bo}_2$ for all the liquids used in our experiments. 
The color encodes the Ohnesorge number, $\mathrm{Oh} = \mu_2 / \sqrt{\rho_2 \Spread w_0}$. 
The experiments nearly collapse onto a master curve of the form given in Eq.~\eqref{eq:scaling_dimensionless} with a fitted constant $C=8$ (solid line). 
The dotted line, corresponding to $C \, \left(\mathrm{Ca} \, \mathrm{Bo}_2\right)^{1/2} \sim 1$, qualitatively marks the transition from limiting behaviors i) to ii).
We find the collapse of the data quite remarkable, taking into account the simplifications of our model.
First, the Ohnesorge number varies more than two orders of magnitude between the different liquids. 
Second, when changing liquids, not only does the Ohnesorge number change but also, to a lesser extent, the ligament wetting properties on water (\textit{i.e.}, Neumann angles) and the Bond number $\mathrm{Bo}_1 = w_0^2/\ell_{c1}^2$.
In regime (I-i), the dimensionless data lies below the prediction of Eq.~\eqref{eq:scaling_dimensionless} and is slightly more scattered. This is to be expected since the precise value of the ratio $\psi = \lim_{\mathrm{Ca}\,\mathrm{Bo}_2\rightarrow 0} V/V_c$ must depend on the details of the geometry of the feeding ligament which, in turn, depend on the physical properties of the liquids. This is also true for the classical dripping faucet, where $\psi$ can be as low as 0.6 \cite{Wilkinson1972}. For this reason, we do not leave $\psi$ as a free parameter in Eq.~\eqref{eq:scaling_dimensionless}, where it is assumed to be 1. Although the agreement with the experimental data would be quantitatively better, a single value of $\psi$ should not formally be valid for all liquids. Moreover, allowing Eq.~\eqref{eq:scaling_dimensionless} to have two fitting parameters would conceal the good job that this equation does at fitting the experimental data over nearly four decades in the parameter $\mathrm{Ca}\,\mathrm{Bo}_2$. Instead, the reasonable agreement obtained with one fitting parameter shows the robustness of the underlying physical mechanisms.
%
\section*{No-pinch-off regime}
%
Increasing further the flow rate and viscosity of liquid 2, experiments show the emergence of another regime (denoted (II) in Fig.~\ref{fig:regime_overview}), in which the hanging ligament never breaks. 
Movie S4, corresponding to RTM30 and $Q = 20~\micro\liter\per\minute$, illustrates the transition towards pinch-off suppression: the ligament wobbles several times before eventually breaking.
This delayed yet still periodic pinch-off translates into a very large lens volume (point marked with an asterisk in Fig.~\ref{fig:Vadim_vs_Ca}), which stands out of the experimental trend.
Increasing further the flow rate (RTM30 and $Q = 50~\micro\liter\per\minute$), pinch-off is totally suppressed (gray area in Fig.~\ref{fig:Vadim_vs_Ca}(b)).
As illustrated in Movie S5, the ligament then continuously feeds a floating puddle of liquid 2 as long as injection continues.

We hypothesize the reason for the occurrence of regime (II) is that the water meniscus is curved, a fact that we ignored so far for the description of the periodic dripping regime (I).
As a consequence of this curvature, the effective gravity felt by the ligament (\textit{i.e.} the driving force promoting pinch-off) decreases as it slides down.
If pinch-off has not occurred by the time the bath surface becomes horizontal and the driving force vanishes, then it never will. 
More quantitatively, the condition for the transition from regime (I) to (II) to happen is that the length $\ell_{po}$ at which the ligament would break exceeds the water meniscus length $\ell_{c1}$.
We can estimate the pinch-off distance as $\ell_{po} \sim U\,t_{po}$, with $t_{po} \sim 6\pi\mu_2 w_0 / \Spread$ the time taken by the stretched ligament to destabilize in the absence of gravity (see section~\ref{apdx:ligament_modelling} in Supp. Mat.).
Thus, the condition $\ell_{c1} \sim \ell_{po}$ for the onset of regime (II) is expressed as $6\pi\,\mathrm{Ca}\,\mathrm{Bo}_1^{1/2} \sim 1$ or equivalently, $\mathrm{Ca}\,\mathrm{Bo}_2 \sim  \mathrm{Bo}_2/ 6\pi\mathrm{Bo}_1^{1/2} \approx 0.16$, roughly the same for all liquids.
This threshold, plotted as a dashed line in Fig.~\ref{fig:Vadim_vs_Ca}(b), is consistent with the experimental data: $6\pi\,\mathrm{Ca}\,\mathrm{Bo}_1^{1/2} = 0.45$ for the last point of periodic pinch-off (RTM30 ; $Q = 20~\micro\liter\per\minute$), and $6\pi\,\mathrm{Ca}\,\mathrm{Bo}_1^{1/2} = 1.13$ for the first point of continuous feeding (RTM30 ; $Q = 50~\micro\liter\per\minute$).
%
\section*{Conclusion} \label{sec:conclusion}
%
In summary, we explore the physics of forced liquid dripping along an immiscible liquid-gas interface, in partial wetting conditions.
This three-phase configuration, dubbed ``interfacial dripping faucet'', exhibits a periodic pinch-off regime in a wide range of control parameters, leading to the generation of monodisperse liquid lenses.
Both the critical volume and the dripping period are found to follow analogous laws to the free, axisymmetric dripping faucet \cite{Wilson1988,Eggers1994}, but with a key difference: the surface tension of the injected liquid must be replaced by the spreading coefficient, $\Spread$. 
This accounts for the fact that the surface tension of the water-air interface does not contribute to keep a lens confined but, on the contrary, pulls to spread it.
Beyond analogies, the interfacial dripping faucet also reveals an original regime not observed in a free vertical dripping faucet.
For sufficiently large viscosity and flow rate of the dispersed phase, pinch-off is suppressed altogether and the liquid ligament continuously feeds a floating puddle.

A very important control parameter of the interfacial dripping faucet is the geometry of the meniscus, which is itself governed by the plate geometry and wettability.
A steeper meniscus could translate into a delayed no-dripping regime, a thinner ligament (and hence smaller lenses), or even in the appearance of multiple dripping periods or a continuous jetting regime. 
The interfacial dripping faucet opens new perspectives to generate two-dimensional emulsions in the absence of confining walls \cite{Desmond2013, Dollet2015} and, more generally, to study the fundamentals of liquid-liquid-gas dispersions, at play in many industrial and natural settings \cite{Hotrum2004, Venkataramani2020, Purohit2014, Su2006, Rossen2017, Shaw2003, Nissanka2018, Gupta2017}.
%
\section*{Acknowledgments}
The authors acknowledge financial support from Grant No. PID2023-146809OB-I00 funded by MICIU/AEI/10.13039/501100011033 and by ERDF/UE (J. R.-R.), Grant No. PID2020-114945RB-C21 funded by MCIN/AEI/10.13039/501100011033 (J. R.-R. and L. C.), and VICI Grant No. 680-47-632 funded by NWO (V. B. and J. H. S.). This project has received funding from the European Union’s Horizon 2020 research and innovation programme under the Marie Sklodowska-Curie Grant Agreement No. 882429 (L. C.).
%
\bibliography{biblio}
\bibliographystyle{unsrt}
\newpage\clearpage
\appendix

\twocolumn[
\begin{@twocolumnfalse}
    \appendixpage
    \begin{center}
        \line(1,0){340}
    \end{center}
    \vspace{0.5cm}
\end{@twocolumnfalse}]

\section{Materials and methods} \label{apdx:matmet}
\subsection{Liquid properties} 
%
\begin{figure}[b]
    \centering
    \includegraphics[width=0.7\columnwidth]{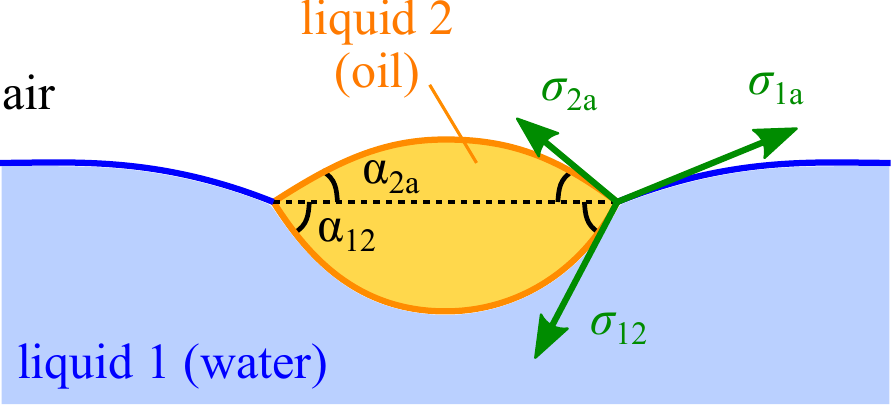}
    \caption{Cross-sectional geometry of a floating lens (or ligament) of liquid 2 on liquid 1.}
    \label{fig:lens_geometry}
\end{figure}
\begin{table*}
\centering
\begin{tabular}{|c|c||c|c|c|c||c|c|c||c|}
\hline Liquid 2 & Refs. & $\rho_2$ & $\mu_2$ & $\sigma_{2a}$ & $\sigma_{12}$ & $\mathcal{S}$ & $\alpha_{12}$ & $\alpha_{2a}$ & $\theta_{2}$ \\
& & ($\kilo\gram\per\cubic\meter$) & ($\mPas$) & ($\mNperm$) & ($\mNperm$) & ($\mNperm$) & ($\degree$) & ($\degree$) & ($\degree$) \\  
\hline
dodecane & \cite{Luning2014, Goebel1997} & 750 & 1.5 & 25 & 53 & -6 & 34 & 15 & 28\\
S6 & \cite{Takamura2012} & 832 & 8.7 & 31 & 54 & -13 & 44 & 23 & 32\\
squalane & \cite{Korotkovskii2012, Comunas2013, Goossens2011} & 810 & 35 & 31 & 56 & -15 & 48 & 24 & 39\\
S60 & \cite{Takamura2012} & 863 & 135 & 31 & 54 & -13 & 44 & 23 & 44\\
S200 & \cite{Takamura2012} & 863 & 563 & 31 & 54 & -13 & 44 & 23 & 49\\
RTM30 & \cite{Takamura2012} & 868 & 7571 & 31 & 54 & -13& 44 & 23 & 65\\ \hline
\end{tabular}
\caption{Physical properties of the fluids used as the dispersed phase (liquid 2), as found in the literature: density $\rho_2$, dynamic viscosity $\mu_2$, surface tension $\sigma_{2a}$ and interfacial tension $\sigma_{12}$ with water. The spreading parameter $\mathcal{S}$, and Neumann angles $\alpha_{2a}$ and $\alpha_{12}$ (calculated from Eqs.~\eqref{eq:Neumann_angles}) are then deduced. The contact angle $\theta_2$ is measured on clean hydrophylized glass. All values are given at room temperature $T \approx 21~\degreecelsius$.}
\label{tab:fluid_properties}
\end{table*}
For all experiments, the liquid substrate (liquid 1) is ultrapure water (Milli-Q) of surface tension $\sigma_{1a} = 72~\mNperm$. 
Six different liquids are used as the dispersed phase (liquid 2): dodecane ($99~\%$, Acros Organics), squalane ($99~\%$, Thermo Fisher), and mineral oils S6, S60, S200 and RTM30 (Paragon Scientific Ltd), all used as received.
Table~\ref{tab:fluid_properties} summarizes relevant properties of those liquids.

As shown by Ravazzoli \textit{et al.} \cite{Ravazzoli2020}, the condition for partial wetting on a liquid substrate (\textit{i.e.} the existence of uncloaked liquid lenses as depicted in figure~\ref{fig:lens_geometry}) is
\begin{equation}
    - 2 \, \mathrm{min}(\sigma_{2a}, \sigma_{12}) < \mathcal{S} < 0,
    \label{eq:lens_condition}
\end{equation}
where $\mathcal{S} = \sigma_{1a} - (\sigma_{12} + \sigma_{2a})$ is the spreading parameter.
For $\sigma_{2a} < \sigma_{12}$, equation~\eqref{eq:lens_condition} boils down to $- 2 \sigma_{2a} < \mathcal{S} < 0$, which is fulfilled for all our liquids, as shown in table~\ref{tab:fluid_properties}.
%
%
\subsection{Solid and liquid substrates}
%
\begin{figure}
    \centering
    \includegraphics[width=0.75\linewidth]{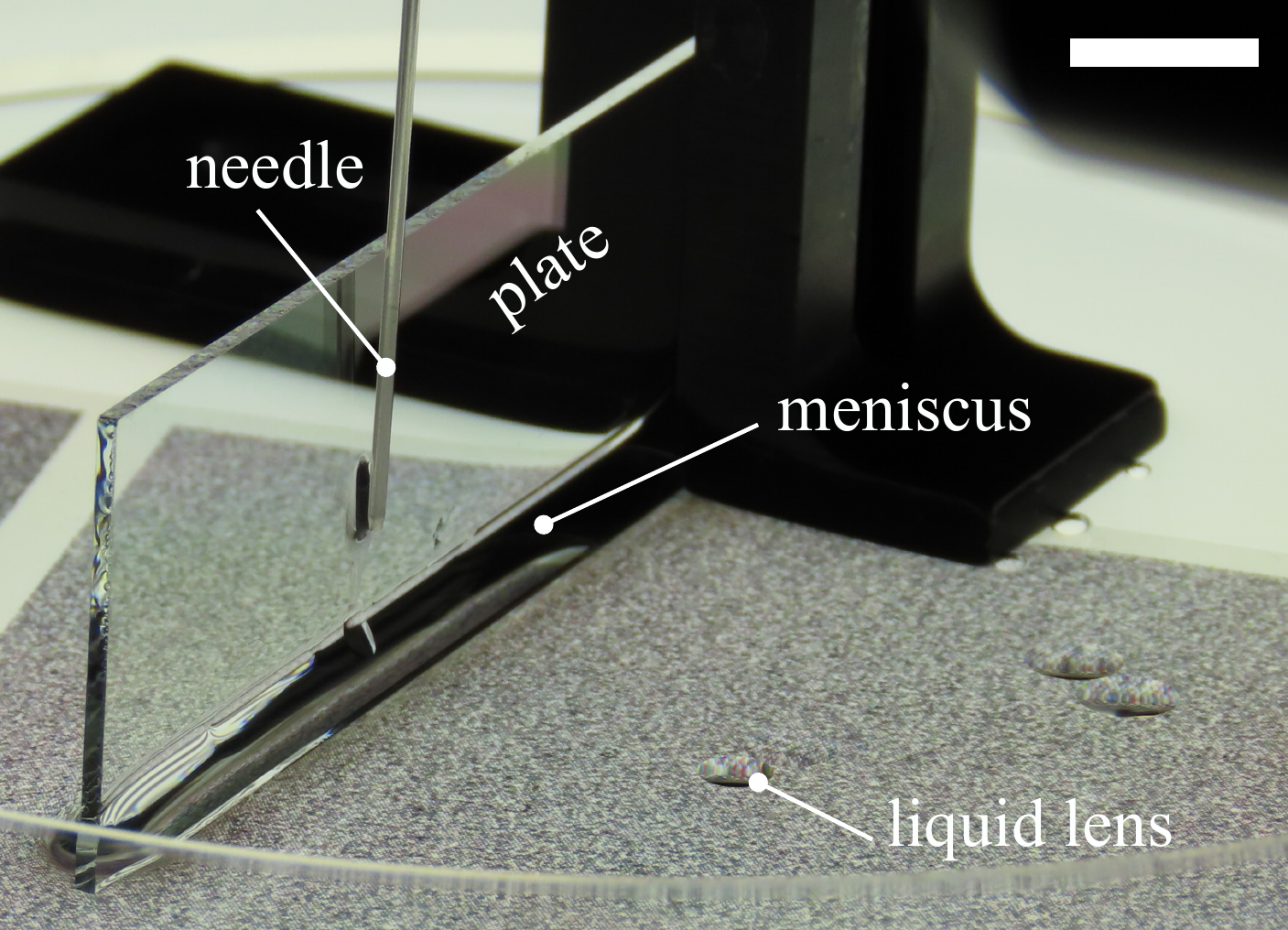}
    \caption{Picture of the experimental configuration, where a random dot pattern underneath the bath is used to ease lens visualization. The scale bar is $1~\centi\meter$.}
    \label{fig:setup_picture}
\end{figure}
An experimental realization of the interfacial dripping faucet geometry is shown in Fig.~\ref{fig:setup_picture}.
In all our experiments, the plate is a microscope glass slide (dimensions $25~\centi\meter \times 75~\centi\meter$), about $1~\milli\meter$ in thickness.
We denote $\theta_1$ (respectively $\theta_2$) the contact angle of liquid 1 (resp. liquid 2) on the plate.

Before use, glass slides are cleaned for $15$ minutes in acetone in an ultrasound bath, then rinsed with ethanol, ultrapure water and isopropanol, and then dried with an air gun.
Finally, the clean slide is hydrophilized in a plasma cleaner to make sure that it is perfectly wetted by water, \textit{i.e.} $\theta_1 = 0 \degree$.
The contact angles $\theta_2$ measured for oils on this substrate are presented in table~\ref{tab:fluid_properties}. The main effect of $\theta_2$ is to modify the width of the oil rivulet dripping on the plate, and in particular its value $w_0$ at the contact line with water (see figure~\ref{fig:w0_vs_Q}).
We also performed a control experiment with a clean non-hydrophilized glass slide (\textit{i.e.} having undergone the above cleaning procedure, but not the hydrophilization step), with water contact angle $\theta_1 \approx 15\degree$.
The dripping period showed no appreciable difference compared to the hydrophilic case.

This plate is held vertically by two right-angle brackets (Thorlabs) inside a disposable Petri dish (diameter $150~\milli\meter$, polystyrene).
The holders are previously cleaned using the same procedure as for the glass slides.
Ultrapure water is then gently poured into the Petri dish, forming a meniscus against the plate, until the water bath reaches a depth of at least $5~\milli\meter$.
%
\subsection{Injection system}
%
The dispersed phase (liquid 2) is loaded in a $1~\milli\liter$ glass syringe (Hamilton), connected to a stainless steel needle (Hamilton) by PEEK tubings and connectors.
Liquid 2 is steadily injected in this system by a syringe pump (Harvard Apparatus), delivering a constant flow rate $Q$ in the range $Q = 1 - 500~\micro\liter\per\minute$.
In all the experiments but one, the needle gauge is G21 ($514~\micro\meter$ inner diameter, $819~\micro\meter$ outer diameter).
We also perform an additional experiment with a G27 needle ($210~\micro\meter$ inner diameter, $413~\micro\meter$ outer diameter).
For a given liquid, the dripping period did not change substantially when changing the needle gauge from G21 to G27.
The needle is held in place vertically, leaning against the plate, with its tip at a distance $\Delta z \approx 6-8~\milli\meter$ above the surface of the undeformed bath (see Fig.~\ref{fig:imaging_processing}a).

When changing to a different dispersed phase, the whole injection system (syringe, tubing, fittings, needle) is flushed with chloroform and then acetone. 
The syringe, fittings and needle are then dismounted and soaked in chloroform for about 10 minutes, along with the right-angle holders.
These components are subsequently sonicated in acetone for 15 minutes and finally rinsed with ethanol, ultrapure water and isopropanol, before being dried with an air gun.
When the phase to be cleaned is dodecane, chloroform is simply substituted by acetone.
%
\subsection{Imaging system}
%
\begin{figure}
    \centering
    \includegraphics[width=\linewidth]{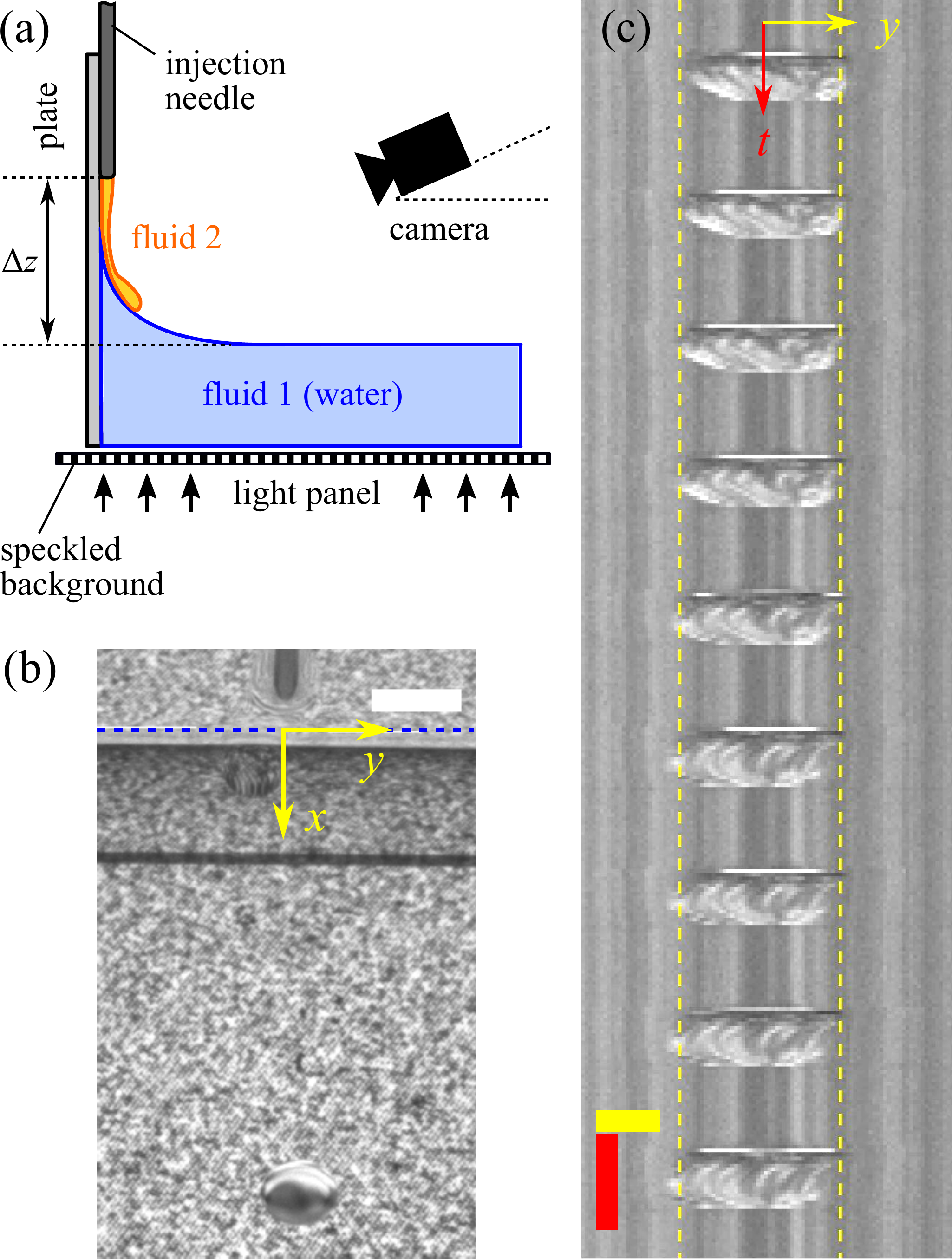}
    \caption{(a) Sketch of the imaging configuration of the interfacial dripping faucet. --- (b) Snapshot of a movie corresponding to a dodecane experiment, with $Q = 100~\micro\liter\per\minute$. The blue dashed line marks the location of the contact line. At the bottom of the snapshot we see the previously detached lens. The scale bar is $5~\milli\meter$. --- (c) Zoomed time-space diagram at the contact line (blue dashed line in panel (b)). The yellow dashed lines mark the maximum width $w_0$ of the ligament at the contact line. The $y$ scale bar (yellow) is $1~\milli\meter$, while the time scale bar (red) is $1~\second$.}
    \label{fig:imaging_processing}
\end{figure}
The interfacial dripping faucet is illuminated from below with a uniform white light panel, as sketched in Fig.~\ref{fig:imaging_processing}a.
During dripping, a camera (Ximea) records a front view at a frame rate between $6$ and $60$ fps, depending on the flow conditions.

The optical index index contrast between liquids 1 and 2 is typically small.
To facilitate the visualization and automatic detection of lenses, we place a speckled background under the water bath (see picture in Fig.~\ref{fig:setup_picture}).
It consists of a random dot pattern with black area fraction of 50\%, laser-printed at 1200 dpi on a transparent plastic sheet.
Before injection of liquid 2 starts, we take a picture of this background that will serve as a reference for image processing (see section~\ref{apdx:image_processing}).

For adequate visualization of the curved meniscus, the camera is inclined at an angle of about $30\degree$ with respect to the horizontal direction.
This introduces a perspective effect that has to be subsequently corrected. 
To do so, we acquire an image of a horizontal, square calibration grid, in the absence of the water-filled bath.
This calibration snapshot will be used in the image processing to compute the transformation matrix that would make the angles between the grid lines right (see section~\ref{apdx:image_processing}).

For a reduced set of experiments, we also visualize the pinch-off dynamics of the dripping ligament with a high-speed camera (Photron Nova S12) equipped with a microscopic lens (Navitar x12 zoom lens).
The frame rate is set between $15$ and $50$ kfps, depending on the flow parameters.
Examples of resulting image sequences are presented in Fig.2b and c of the main text.
%
\section{Image processing} \label{apdx:image_processing}
%
In this section, we explain how the acquired videos are processed to measure two magnitudes of interest: the dripping interval, $\Delta t$, and the dripping ligament width at the contact line, $w_0$.
%
\subsection{Dripping interval $\Delta t$}
%
\begin{figure}
    \centering
    \includegraphics[width=\columnwidth]{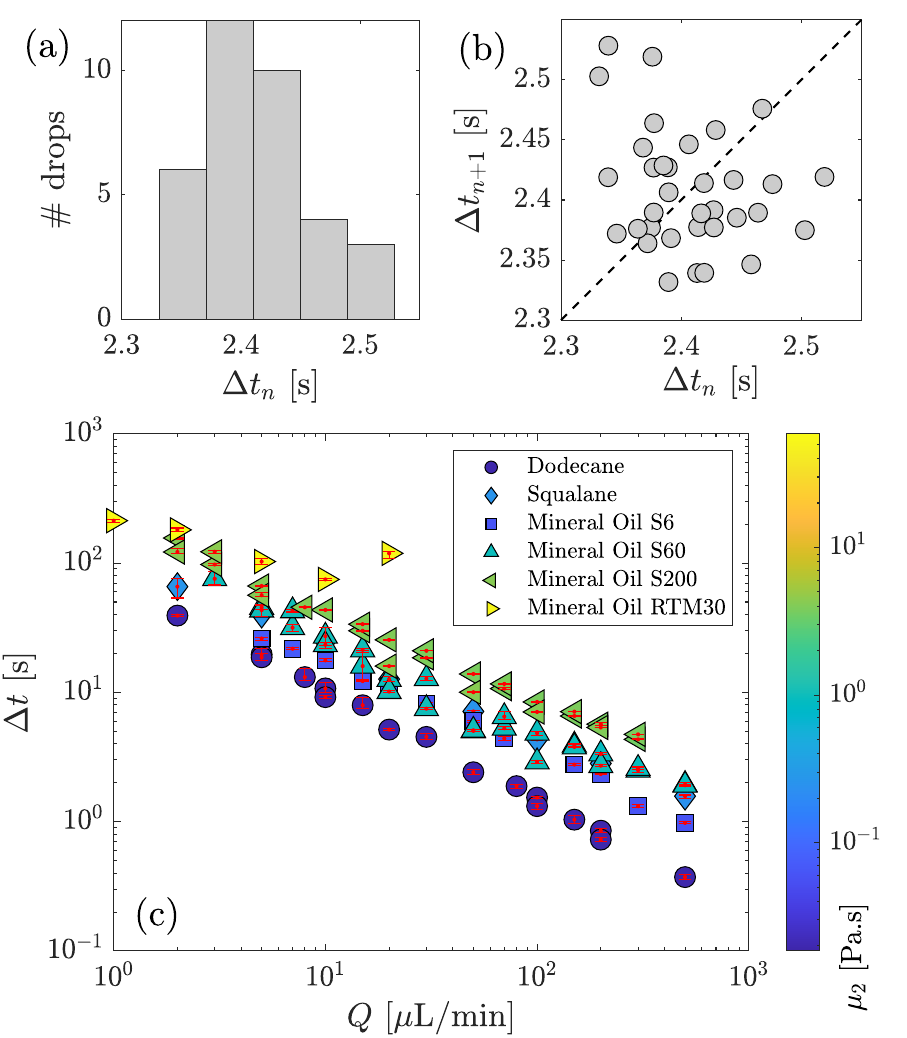}
    \caption{(a) Dripping period $\Delta t$ as a function of the flow rate $Q$ for different injected liquids. Error bars (in red) denote the minimum and the maximum value in the series. Markers' colors correspond to the dynamic viscosity $\mu_2$. (b) Histogram of dripping intervals $\Delta t_n$ for a given experimental session (dodecane, $Q=50~\micro\liter\per\minute$). (c) Time-return plot of the dripping intervals for the same session.}
    \label{fig:dt_vs_Q}
\end{figure}
We detect the time evolution of the position of the liquid lenses, once detached from the feeding rivulet, by digital processing of images like that shown in Fig. \ref{fig:imaging_processing}b. 
The algorithm, implemented by a house-made code written in MATLAB$^\copyright$ 2020a, consists in applying the following steps to all the images of a video:
\begin{enumerate}
    \item The reference speckled background image is subtracted.
    \item The resulting image is corrected for perspective effects using the transformation matrix obtained from the image of the calibrated grid.
    \item The transformed image is binarized using a gray-level threshold.
    \item A morphological closing operation is applied to the binary image.
    \item Objects larger than a threshold size are detected, in particular recording the location of their centers as a function of time.
\end{enumerate}
After the images are processed, a home-made tracking algorithm is applied to detect the trajectories of individual lenses. Finally, we define the dripping interval of lens number $n$, $\Delta t_n$, as the interval between the time when the lens crosses a given position $x$ (perpendicular to the plate), and when the next lens crosses the same position (Fig. 1c in main text). Then, we compute the dripping period as the mean dripping interval
\begin{equation}
    \Delta t = \frac{1}{N}\sum\limits_{n=1}^N \Delta t_n,
\end{equation}
where $N$ is the total number of lenses. 
The dripping periods obtained applying this procedure to all the movies are shown in Fig.~\ref{fig:dt_vs_Q}a.

To quantify the fluctuations of $\Delta t_n$ around $\Delta t$, we plot in Fig.~\ref{fig:dt_vs_Q}b the histogram of all the dripping intervals measured for a given experiment (here dodecane, $Q=50~\micro\liter\per\minute$). 
The typical variability in dripping interval, defined as the standard deviation of $\Delta t_n$ divided by its average value, is of the order of $5$ to $10$\%.

Period doubling or chaotic dripping regimes have been reported in the classical dripping faucet when $\Delta t_n$ is of the order of or smaller than the typical recoiling time of the filament after pinch-off \cite{Clanet1999, Coullet2005}. 
To assess the periodicity of the interfacial dripping faucet, we compute the time-return plot of the dripping interval in Fig.~\ref{fig:dt_vs_Q}c.
In case of period doubling, for example, one would observe two clusters of points symmetric with respect to the bisector line \cite{Subramani2006}.
Here, the data is scattered around $\Delta t_{n+1} = \Delta t_n$ (dashed line) with no particular trend, suggesting that successive dripping events are uncorrelated.
This feature is observed across the whole parameter range we investigated, showing that the dripping period $\Delta t$ is well-defined for the interfacial dripping faucet.
%
\subsection{Ligament width $w_0$}
%
\begin{figure}
    \centering
    \includegraphics[width=\columnwidth]{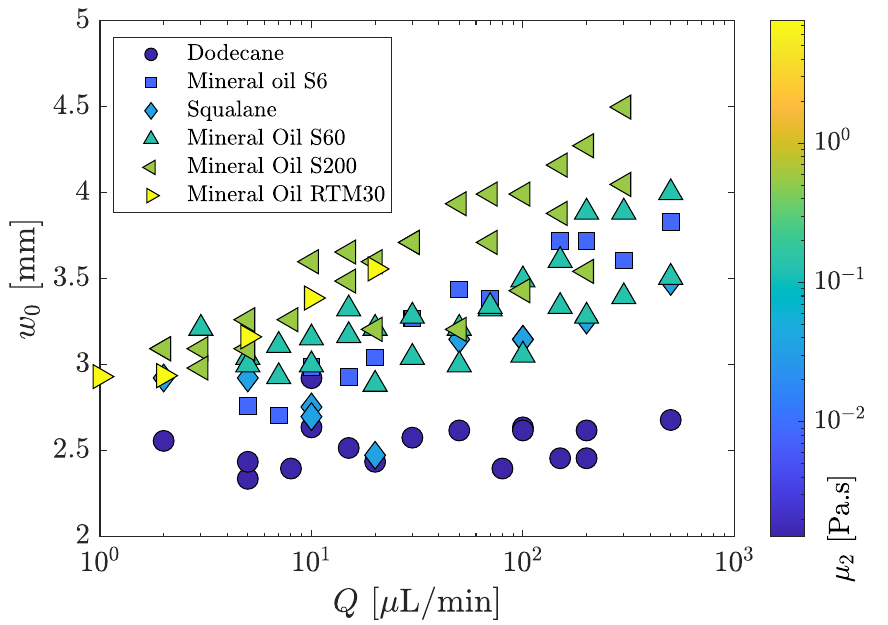}
    \caption{Ligament maximum width at the contact line, $w_0$, as a function of the flow rate $Q$ for different injected liquids. Colors correspond to the dynamic viscosity $\mu_2$.}
    \label{fig:w0_vs_Q}
\end{figure}
The main length scale we use to describe the production of liquid lenses is the dripping ligament width $w_0$ at the contact line.
This magnitude is not trivial to measure directly from the images, especially when liquid 2 is not very viscous.
In this parameter range, the injected liquid recoils right after pinch-off, and no ligament can therefore be seen at the contact line during part of the dripping process. 
This is for instance the case in the example shown in Fig.~\ref{fig:imaging_processing}b. 

Given this limitation, the \emph{maximum} ligament width at the contact line is the only observable that can be consistently obtained for all the experimental conditions.
We will therefore use it as the definition of $w_0$.
To measure it, we extract the slice (\textit{i.e.} pixel line) corresponding to the contact line in each frame of the video.
All those slices make up a synthetic time-space diagram (Fig.~\ref{fig:imaging_processing}c), where the $m$-th line contains the slice from the $m$-th image.
An automated computer algorithm is then used to detect the maximum width of the ligament in this synthetic image (dashed yellow lines in Fig.~\ref{fig:imaging_processing}c).
We show in Fig.~\ref{fig:w0_vs_Q} the variation of $w_0$ as a function of the flow rate for various viscosities of the disperse phase.
%
\section{Modelling the floating ligament} \label{apdx:ligament_modelling}
\subsection{Geometry} 
%
The cross-sectional geometry of a floating lens (or ligament) of liquid 2 on liquid 1 is sketched in Fig.~\ref{fig:lens_geometry}. The Neumann angles $\alpha_{2a}$ and $\alpha_{12}$ formed by the top (liquid 2/air) and bottom (liquid 1/liquid 2) interfaces of a liquid lens or ligament are computed from the force balance at the triple contact line liquid 1/liquid 2/air \cite{Ravazzoli2020}:
\begin{eqnarray}
    \sigma_{2a}\sin\alpha_{2a} - \sigma_{12}\sin\alpha_{12} & = & 0,\\
    \sigma_{2a}\cos\alpha_{2a} + \sigma_{12}\cos\alpha_{12} & = & \sigma_{1a}.
    \label{eq:Neumann_angles}
\end{eqnarray}
These angles are used to obtain the cross-sectional area $A_0$ of the ligament downstream the plate-water-air contact line.
%
The cross-section is assumed to be bounded by two circular arcs that form Neumann angles $\alpha_{2a}$ and $\alpha_{12}$ with respect to a horizontal bath surface (Fig.~\ref{fig:lens_geometry}).
Knowing the ligament width $w_0$, the cross-sectional area is deduced as
\begin{equation}
    A_0 = \frac{w_0^2}{2}\left(\frac{\alpha_{12}}{\sin^2\alpha_{12}} - \frac{1}{\tan\alpha_{12}} + \frac{\alpha_{2a}}{\sin^2\alpha_{2a}} - \frac{1}{\tan\alpha_{2a}}\right).
\end{equation}
%
\subsection{Force balance on the hanging ligament in the quasi-static dripping regime}
%
\begin{figure*}
        \centering
        \includegraphics[width=0.75\linewidth]{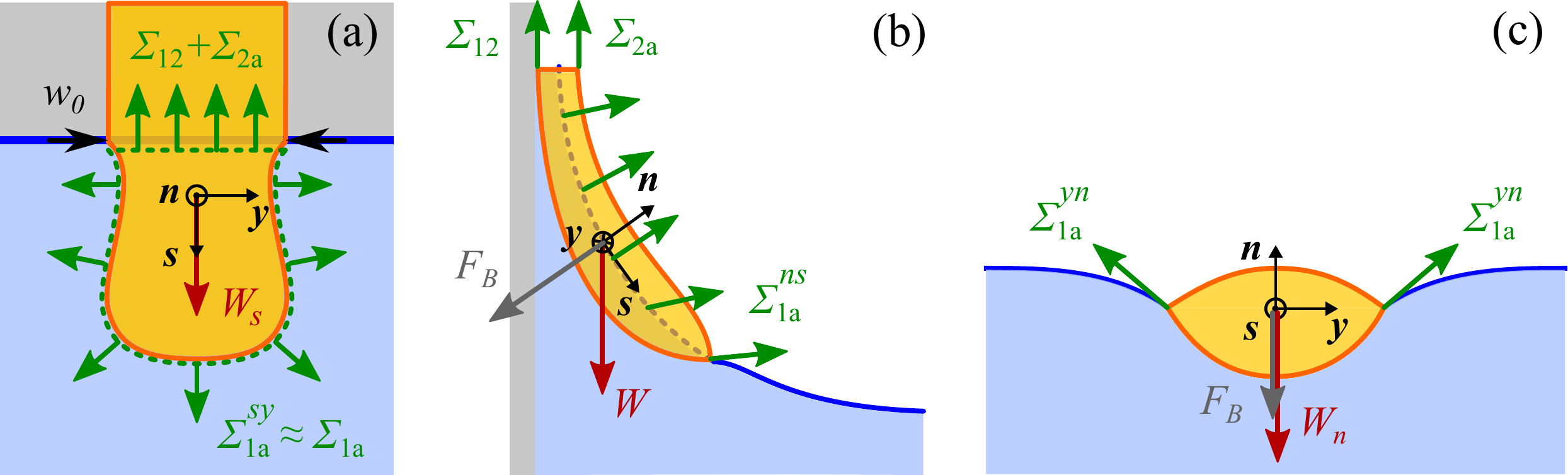}
        \caption{Sketch of three views of the model geometry we use to compute the force balance on the hanging liquid ligament. (a) ``Front'' view, perpendicular to the water-air free surface. (b) Cut with a vertical plane perpendicular to the plate. (c) Cut perpendicular to the main axis of the ligament.}
        \label{fig:force_sketch}
\end{figure*}
In this subsection we provide further details on the derivation and hypotheses behind the force balance that we use in the main manuscript to estimate the maximum lens volume $V_c$ that can hang from the feeding rivulet in the quasi-static dripping regime. 
To help visualizing the different forces, Fig.~\ref{fig:force_sketch} displays three views of the hanging ligament: (a) a planar development of the curved water-air meniscus, (b) a cut through the ligament's mid-plane, and (c) a cut perpendicular to the ligament central axis. 
We define the following coordinate system: a streamwise coordinate $s$ that runs parallel to the undisturbed water-air interface, at the intersection with the ligament symmetry plane ; a normal coordinate, $n$, contained also in the symmetry plane, but perpendicular to the meniscus ; the $y$ coordinate, perpendicular to both $s$ and $n$. 
In panel (a) we show with a green dashed line the projection on the $s-y$ plane of the control volume on which the force balance is performed.

To compute the force balance, we introduce two main simplifications:
\begin{itemize}
    \item We treat the ligament as if it had a vanishingly small thickness -- although for the sake of clarity it is depicted with a finite thickness in Fig.~\ref{fig:force_sketch}. 
    This is a reasonable assumption as long as the Neumann angles are rather small (see table~\ref{tab:fluid_properties}).
    \item Second, we assume that the plane $s-y$ is actually vertical. Strictly speaking, this is true only at the triple contact line plate/water/air, owing to the hydrophilic treatment applied to the plate such that $\theta_1 = 0$. 
    However, as long as the ligament length is smaller than the typical size of the meniscus (given by the capillary length $\ell_{c1}$), it will evolve in a region where the $s-y$ plane is close to vertical.
\end{itemize}
This second assumption has two important consequences. First, the ligament is hanging mostly parallel to gravity, so that its weight $W$ can be estimated as $\rho_2 g V_c$.
Second, the buoyancy force acting on the ligament, $F_B$, is nearly horizontal (see Fig.~\ref{fig:force_sketch}b). 
Indeed, buoyancy is the resultant between the ambient pressure acting on the liquid 2/air interface, and the pressure inside the meniscus, which acts on the liquid 1/liquid 2 interface.
Assuming the ligament is flat, the resultant force is then locally parallel to the direction $n$, that is, approximately horizontal. 
This means that buoyancy is expected to have only a small contribution in the vertical force balance. 
Moreover, since the pressure inside the water meniscus is smaller than the ambient one, buoyancy pushes the ligament towards the water bath.

In those conditions, the weight of the ligament is counterbalanced mainly by capillary forces. 
The liquid 2/air and liquid 1/liquid 2 interfaces pull the ligament up, with a total force $\Sigma_{2a} + \Sigma_{12} = w_0\left(\sigma_{2a} + \sigma_{12}\right)$. To obtain this expression we have again made use of the hypothesis that the ligament is flat.
Another capillary force arises from the water/air surface, pulling on the water/liquid 2/air triple contact line of the ligament with a surface tension $\sigma_{1a}$ (Fig. \ref{fig:force_sketch}a). 
Using again the assumption that the ligament is flat and that its presence does not deform the (nearly vertical) water/air interface, the resultant of this distributed surface tension is $\Sigma_{1a} = w_0 \sigma_{1a}$, pulling vertically downwards.
Note that the presence of the ligament does induce some deformation on the water/air surface, which leads to the appearance of a small projection of the surface tension force $\Sigma_{1a}$ along the normal ($n$) direction, as sketched in Fig. \ref{fig:force_sketch}c. 
This projection must be in equilibrium with the buoyancy force and a small projection of the weight onto the $n$ direction. 
However, this does not affect the critical volume $V_c$, which arises from the vertical force balance.

In summary, putting all these ingredients together, the weight of the hanging ligament must be supported by the forces associated to the two tensions connecting it to the feeding rivulet, $\Sigma_{12}$ and $\Sigma_{2a}$, minus the pull exerted by the bath's free surface, $\Sigma_{1a}$. Mathematically,
\begin{equation}
    \rho_2 g V_c = - w_0 \left(\sigma_{1a} - \sigma_{2a} - \sigma_{12}\right) = w_0 \Spread.
\end{equation}
As pointed out in the main manuscript, this critical volume $V_c$ slightly overestimates the actual detached volume $V$, as also happens in the classical dripping faucet. 
Once the volume $V_c$ is reached, the ligament becomes unstable and starts to pinch, but the exact detached volume depends on the particularities of the pinch-off process, as some liquid usually remains attached to the feeding ligament \cite{Wilkinson1972}. 
Still, our estimation of $V_c$ gives a reasonable approximation of the lens volume measured experimentally in the quasi-static dripping regime. 
At the same time, it avoids dealing with the details of the very complex geometry of the hanging ligament.
%
\subsection{Analytical model of detached lens volume based on Wilson (1988)}
%
\begin{figure}
\centering
\includegraphics[width=\columnwidth]{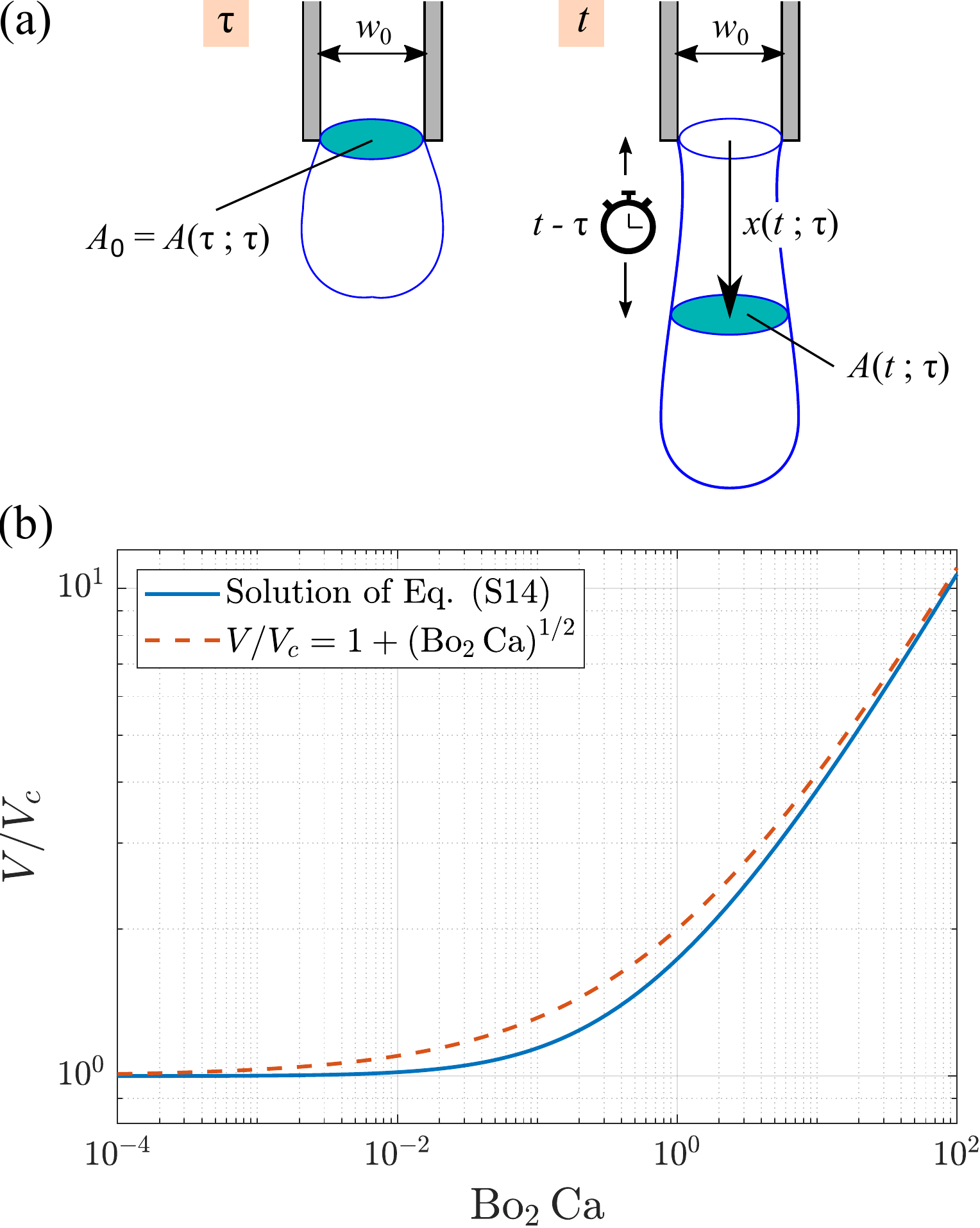}
\caption{(a) Sketch of the Lagrangian coordinates used in Wilson's analysis. (b) Comparison between the solution of Eq. (\ref{eq:VVc_vs_Bo2Ca}) and the approximate asymptotic expression $V/V_c = 1 + \left(\mathrm{Bo_2} \, \mathrm{Ca}\right)^{1/2}$.}
    \label{fig:comparison_Wilson}
\end{figure}
Eq. (1) of the main manuscript, which we rewrite here for convenience and future reference, was shown to reproduce reasonably well the volume of detached liquid lenses measured in the periodic dripping regime: 
\begin{equation}
    \frac{V}{V_\mathrm{c}} = 1 + C \, \left(\mathrm{Ca} \, \mathrm{Bo}_2\right)^{1/2}.
    \label{eq:volume_approximation}
\end{equation}
This expression was built by summing the time scales of the two processes that must be occur sequentially for a lens to detach (liquid accumulation and ligament stretching).
Eg.~\eqref{eq:volume_approximation} represents a linear interpolation between the two regimes in which either one process or the other sets the dripping period.
The objective of this section is to assess how quantitative is this approximation.
To do so, we adapt the rigorous model of Wilson \cite{Wilson1988} for the vertical, axisymmetric dripping faucet to describe our ``interfacial dripping faucet'' configuration.

Consider a vertical liquid jet flowing out of a nozzle of diameter $w_0$ and with a cross sectional area $A_0$. 
Wilson's geometry (pictured in Fig.~\ref{fig:comparison_Wilson}a) is axisymmetric but we will not make this assumption, so that the model may be applied to more general configurations.
We suppose that the velocity field is one-directional to leading order, directed along the streamwise coordinate $x$ and homogeneous at each cross section of the ligament. 
We will use a Lagrangian formulation to determine the viscous stretching time of the jet. 
In this formulation, each flow cross-section is labelled with a Lagrangian time, $\tau$, which corresponds to the instant in time when the fluid at this section was injected (Fig.~\ref{fig:comparison_Wilson}a). 
We define $A(t; \tau)$ as the cross-sectional area, at a time $t$, of the fluid section injected at an earlier time $\tau$. 
Consistently, at time $t = \tau$, when this fluid section was emerging from the nozzle, we had $A(\tau; \tau) = A_0$. 
Note that in this formulation the time variable is $t$, whereas $\tau$ is a parameter that labels each fluid section (in Fig.~\ref{fig:comparison_Wilson}a, the one shaded in blue).
Finally, we suppose that the viscosity of the liquid is large enough such that inertia is negligible. 

At each cross-section $\tau$ the weight of the liquid injected up to this point, $\rho_2 g Q \tau$, is balanced by the capillary forces and by the force exerted by Trouton elongational stresses acting on the section.
The latter is simply $3 \mu_2 \partial A / \partial t$, while the former depends on the exact shape of the cross-section.
We introduce the dimensionless parameter 
\begin{equation}
m = \frac{P(t; \tau)}{\sqrt{A(t; \tau)}},
    \label{ed:def_m}
\end{equation}
where $P(t; \tau)$ is the perimeter of the cross-section -- on which capillary forces act -- and $A(t; \tau)$ its area.
In the axisymmetric geometry of Wilson $m = 2\sqrt{\pi}$ but here we will simply make the assumption that $m$ does not depend on $t$ nor $\tau$.
This model may be used to describe the interfacial dripping faucet provided that shear stresses exerted by the liquid bath on the hanging ligament can be neglected.
In that case, the geometrical parameter $m$ is a function of the Neumann angles, and thus of the liquid properties.
Finally replacing surface tension by the modulus of the spreading coefficient $\Spread$, the force balance on a given cross-section reads:
\begin{equation}
    -3\mu_2\frac{\partial A}{\partial t} + m \Spread A^{1/2} = \rho_2 g Q \tau.
    \label{eq:Reb_equilibrium_dimensional}
\end{equation}

We introduce the following dimensionless variables: $\hat{A} = A/A_0$, and $\hat{t} = Qt / (m \Spread A_0^{1/2}/ \rho_2 g)$.
Note that the dimensionless time $\hat{t}$ can be interpreted as the ratio between the liquid volume injected up to time $t$, namely $Qt$, and the maximum volume that capillary forces can sustain against the liquid weight, $V_c = m \Spread A_0^{1/2} / \rho_2 g$.
The Lagrangian time $\hat{\tau}$ is non-dimensionalized the same way as $\hat{t}$.
Using those dimensionless variables, Eq. (\ref{eq:Reb_equilibrium_dimensional}) can be recast as
\begin{equation}
    -K\,\mathrm{Bo_2}\,\mathrm{Ca}\frac{\partial \hat{A}}{\partial \hat{t}} + \hat{A}^{1/2} = \hat{\tau},
    \label{eq:Reb_equilibrium}
\end{equation}
where $\mathrm{Bo_2}$ and $\mathrm{Ca}$ are respectively the Bond and capillary numbers defined in the manuscript.
The parameter $K$ is defined as $K = 3F/m^2$, where $F = A_0 / w_0^2$ is another geometrical parameter.
Equation (\ref{eq:Reb_equilibrium}) can be integrated for a given fluid section between the time at which it was injected, $\hat{\tau}$, and the time at which it would pinch-off, $\hat{t}_{po}$:
\begin{equation}
    K\,\mathrm{Bo_2}\,\mathrm{Ca} \int_1^0 \frac{\mathrm{d}\hat{A}}{\hat{A}^{1/2}-\hat{\tau}} = \int_{\hat{\tau}}^{\hat{t}_{po}} \mathrm{d}\hat{t},
\end{equation}
resulting in
\begin{equation}
-2K\,\mathrm{Bo_2}\,\mathrm{Ca} \left(1 + \hat{\tau}\,\log\left(1 - \frac{1}{\hat{\tau}}\right)\right) = 
\hat{t}_{po}-\hat{\tau}.
\label{eq:adim_tpo_vs_tau}
\end{equation}
Equation \eqref{eq:adim_tpo_vs_tau} predicts the time $\hat{t}_{po}$ at which a liquid section injected at time $\hat{\tau}$ pinches off.
The fluid section that pinches off first, and thus makes the lens split from the ligament, corresponds to the minimum value of $\hat{t}_{po}(\hat{\tau})$. 
To find this fluid section, denoted $\hat{\tau}_{po}$, we differentiate equation \eqref{eq:adim_tpo_vs_tau} with respect to $\hat{\tau}$ and impose $\partial \hat{t}_{po}/\partial \hat{\tau} = 0$, yielding
\begin{equation}
    \log\left(1 - \frac{1}{\hat{\tau}_{po}}\right) + \frac{1}{\hat{\tau}_{po}-1} = \frac{1}{2K\mathrm{Bo_2}\,\mathrm{Ca}}.
\end{equation}
Let us recall that, with the non-dimensionalization used, $\hat{\tau}_{po}$ represents the volume detached at pinch-off, divided by the critical volume, \textit{i.e.} $\hat{\tau}_{po} = V/V_c$, hence
\begin{equation}
    \log\left(1 - \frac{1}{V/V_c}\right) + \frac{1}{V/V_c-1} = \frac{1}{2K\mathrm{Bo_2}\,\mathrm{Ca}}.
    \label{eq:VVc_vs_Bo2Ca}
\end{equation}
Eq.~\eqref{eq:VVc_vs_Bo2Ca} is an (implicit) analytical expression giving the volume of drops detaching from a laminar jet of arbitrary cross-section.
For the axisymmetric case described by Wilson, $K$ is a universal constant equal to $3\sqrt{\pi}/8$ but, in the interfacial dripping faucet configuration, $K$ depends on the liquid properties through the Neumann angles.

For small values of the parameter $\mathrm{Bo}_2\,\mathrm{Ca}$, Eq.~\eqref{eq:VVc_vs_Bo2Ca} predicts $V/V_c \approx 1$.
Conversely, in the limit where $\mathrm{Bo}_2\,\mathrm{Ca}$ is large, we get $V/V_c \approx (K \mathrm{Bo}_2\,\mathrm{Ca})^{1/2}$ from a Taylor expansion in $(V/V_c)^{-1} \ll 1$.
Both limits are consistent with the scaling arguments developed in the main manuscript.
In Fig.~\ref{fig:comparison_Wilson}b, we compare the uniform approximation between those two asymptotic limits (Eq.~\eqref{eq:volume_approximation} with $C=1$), to the numerical solution of Eq. (\ref{eq:VVc_vs_Bo2Ca}) with $K=1$.
We can see that Eq.~\eqref{eq:volume_approximation} provides an almost quantitative approximation to the solution of Eq. (\ref{eq:VVc_vs_Bo2Ca}), while being simpler to evaluate and interpret.
%
\subsection{Pinch-off time in the absence of gravity}
%
To estimate the pinch-off time of our floating ligament when it is on the horizontal part of the bath we make the assumption that it can be treated as a free circular viscous cylinder. Doing so, we can apply the calculations done in \S111 of Chandrasekhar \cite{Chandrasekhar1981}. 
Adapting their equation (106) to our notation, the perturbed ligament width is expressed as
\begin{equation}
    w(x, t) = w_0 + \varepsilon_0\,\exp\left[-i (\omega t - k x)\right],
\end{equation}
where $w_0$ is the unperturbed ligament width, $\varepsilon_0$ the perturbation amplitude, $k = 2\pi / \lambda$ the angular wave number corresponding to a wavelength $\lambda$ and $\omega$ the angular frequency. 
For very viscous fluids, the imaginary part of the angular frequency is given by (Eq. (161) of ref. \cite{Chandrasekhar1981})
\begin{equation}
    \omega_i = \frac{\Spread}{3\mu_2 w_0}\left[1 - \left(\frac{k w_0}{2}\right)^2\right].
\end{equation}
The fastest-growing mode of the Rayleigh-Plateau instability is the one that maximizes $\omega_i$, which occurs in the limit of very long waves, $k w_0 \ll 1$. 
Denoting the pinch-off time as the growth time of this fastest-growing perturbation, we finally obtain
\begin{equation}
    t_{po} = \frac{2\pi}{\omega_i} = \frac{6\pi \mu_2 w_0}{\Spread}.
\end{equation}
%

\end{document}